\documentclass[twocolumn,nopacs,preprintnumbers,amsmath,amssymb,aps,floatfix]{revtex4-1}
\usepackage{graphicx}
\usepackage{dcolumn}
\usepackage{bm}

\usepackage{graphicx}
\usepackage{color}
\makeatother 
\begin{document}
\title{Memory improves precision of cell sensing in fluctuating environments}
 \author{Gerardo Aquino$^1$}
 \author{ Luke Tweedy$^{1,2}$}
 \author{Doris Heinrich$^{3,4}$}
 \author{Robert G. Endres$^1$}
 \affiliation{$^1$ Department of Life Sciences and Centre for Systems Biology and Bioinformatics, Imperial College, London, SW7 2AZ, United Kingdom,}\affiliation{$^2$Beatson Institute for Cancer Research, Glasgow, G61 1BD, UK,} \affiliation{$^3$ Leiden Institute of Physics, Leiden University, Leiden, The Netherlands}\affiliation{$^4$ Center for NanoScience (CeNS), Ludwig-Maximilians-Universitat, Geschwister-Scholl-Platz 1, 80539 Munich, Germany.}

\newcommand{\gammaa}{\gamma_{\alpha}}
\newcommand{\e}{{\rm e}}
\newcommand{\veps}{\varepsilon}
\newcommand{\hy}{\hat{y}}
\newcommand{\hu}{\hat{u}}
\newcommand{\hx}{\hat{x}}
\newcommand{\hs}{\hat{\sigma}}
\newcommand{\hsp}{\hat{\sigma'}}
\newcommand{\lgl}{\langle}
 \newcommand{\rgl}{\rangle}
\renewcommand{\d}{{\rm d}}
\newcommand{\singlespace}{\renewcommand{\baselinestretch}{1.0}}
\newcommand{\Vh}[1]{\hat{#1}}
\newcommand{\Aa}{A^1_{\epsilon}}
\newcommand{\Ab}{A^{\epsilon}_L}
\newcommand{\Ae}{A_{\epsilon}}
\newcommand{\finn}[1]{\phi^{\pm}_{#1}}
\newcommand{\ea}{e^{-|\alpha|^2}}
\newcommand{\eb}{\frac{e^{-|\alpha|^2} |\alpha|^{2 n}}{n!}}
\newcommand{\ebbb}{\frac{e^{-3|\alpha|^2} |\alpha|^{2 (l+n+m)}}{l!m!n!}}
\newcommand{\ass}{\alpha}
\newcommand{\as}{\alpha^*}
\newcommand{\fb}{\bar{f}}
\newcommand{\gb}{\bar{g}}
\newcommand{\la}{\lambda}
\newcommand{\sz}{\hat{s}_{z}}
\newcommand{\sy}{\hat{s}_y}
\newcommand{\sx}{\hat{s}_x}
\newcommand{\sio}{\hat{\sigma}_0}
\newcommand{\six}{\hat{\sigma}_x}
\newcommand{\siz}{\hat{\sigma}_{z}}
\newcommand{\siy}{\hat{\sigma}_y}
\newcommand{\vhsig}{\vec{hat{\sigma}}}
\newcommand{\hsig}{\hat{\sigma}}
\newcommand{\hI}{\hat{I}}
\newcommand{\hone}{\hat{1}}
\newcommand{\hH}{\hat{H}}
\newcommand{\hU}{\hat{U}}
\newcommand{\hA}{\hat{A}}
\newcommand{\hB}{\hat{B}}
\newcommand{\hC}{\hat{C}}
\newcommand{\hD}{\hat{D}}
\newcommand{\hV}{\hat{V}}
\newcommand{\hW}{\hat{W}}
\newcommand{\hK}{\hat{K}}
\newcommand{\hX}{\hat{X}}
\newcommand{\hM}{\hat{M}}
\newcommand{\hN}{\hat{N}}
\newcommand{\hO}{\hat{O}}
\newcommand{\te}{\theta}
\newcommand{\vze}{\vec{\zeta}}
\newcommand{\vet}{\vec{\eta}}
\newcommand{\vx}{\vec{\xi}}
\newcommand{\vc}{\vec{\chi}}
\newcommand{\hro}{\hat{\rho}}
\newcommand{\vro}{\vec{\rho}}
\newcommand{\hR}{\hat{R}}
\newcommand{\half}{\frac{1}{2}}
\renewcommand{\d}{{\rm d}}
\renewcommand{\top }{ t^{\prime } }
\newcommand{\kp}{k^{\prime}}
\newcommand{\kpp}{{k^{\prime}}^2}
\newcommand{\xb}{\bar{x}}
\newcommand{\sint}{{\rm si}}
\newcommand{\cint}{{\rm ci}}
\newcommand{\de}{\delta}
\newcommand{\ep}{\varepsilon}
\newcommand{\De}{\Delta}
\newcommand{\eps}{\varepsilon}
\newcommand{\si}{\hat{\sigma}}
\newcommand{\om}{\omega}
\newcommand{\tr}{{\rm tr}}
\newcommand{\ha}{\hat{a}}
\newcommand{\gam}{\gamma ^{(0)}}
\newcommand{\pe}{\prime}
\newcommand{\BEQ}{\begin{equation}}
\newcommand{\EEQ}{\end{equation}}
\newcommand{\BEA}{\begin{eqnarray}}
\newcommand{\EEA}{\end{eqnarray}}
\newcommand{\sph}{spin-$\frac{1}{2}$ }
\newcommand{\ad}{\hat{a}^{\dagger}}
\newcommand{\add}{\hat{a}}
\newcommand{\spp}{\hat{\sigma}_+}
\newcommand{\smm}{\hat{\sigma}_-}
\newcommand{\fin}[1]{|\phi^{\pm}_{#1}\rangle}
\newcommand{\finp}[1]{|\phi^{+}_{#1}\rangle}
\newcommand{\finm}[1]{|\phi^{-}_{#1}\rangle}
\newcommand{\lfin}[1]{\langle \phi^{\pm}_{#1}|}
\newcommand{\lfinp}[1]{\langle \phi^{+}_{#1}|}
\newcommand{\lfinm}[1]{\langle \phi^{-}_{#1}|}
\newcommand{\lfinn}[1]{\langle\phi^{\pm}_{#1}|}
\newcommand{\z}{\cal{Z}}
\newcommand{\RI}{\hat{{\cal{R}}}_{0}}
\newcommand{\Rt}{\hat{{\cal{R}}}_{\tau}}
\newcommand{\nn}{\nonumber}




\begin{abstract}
Biological cells are often found to sense their chemical environment near the single-molecule detection limit. Surprisingly, this precision is higher than simple estimates of the fundamental physical limit, hinting towards 
 active sensing strategies. In this work, we analyse the effect of cell memory, e.g. from slow biochemical processes, on the precision of sensing by cell-surface receptors. We derive analytical formulas, which show that memory significantly improves sensing in weakly fluctuating environments. However, surprisingly when memory is adjusted dynamically, the precision is always improved, even in strongly fluctuating environments. \textcolor{black}{In support of} this prediction we quantify the directional biases in chemotactic {\em Dictyostelium discoideum} cells in a flow chamber with alternating chemical gradients. The strong similarities between cell sensing and control engineering suggest universal problem-solving strategies of living matter.
\end{abstract}


\maketitle






The survival and function of cells and organisms crucially depend on precise sensing of the environment \cite{PNR1,PNR2,PNR3}.   When searching for nutrients or avoiding toxins the bacterium {\it Escherichia coli} can detect differences in concentration as low as 3nM \cite{PNR4}, amounting to approximately 3 molecules per cell volume. T cells can detect single copies of foreign antigen \cite{PNR5} to quickly launch an immune response, while axonal growth cones accurately detect very few molecules of guidance cues (e.g. netrins, slits and ephrins) to follow molecular gradients while seeking their synaptic target  \cite{PNR6}. Such high precision appears to be remarkable since sensing and signalling in a cell is affected by many sources of noise  \cite{PNR2,PNR7,PNR8}. However, what level of precision do we expect from theory? 

Take, for instance, {\em Dictyostelium  discoideum}, which is a well-studied organism both in its unicellular (amoeba) and aggregate state \textcolor{black}{(slug) \cite{PNR9,PNR10}. Under starvation these} amoebae are known to chemotax over a wide range of \textcolor{black}{cyclic adenosine monophosphate  (cAMP)} concentrations, ranging from 0.1nM to 10$\mu$M, with corresponding concentration differences of only 1-5\% across the cell length  \cite{PNR11,PNR12}. Surprisingly, estimates of the receptor-occupancy difference between cell front and back (signal) are dwarfed by occupancy fluctuations (noise) \textcolor{black}{\cite{PNR11}}.  Consequently, the chemotactic ability of these amoebae to aggregate during starvation is better than what should be possible theoretically. This is particularly puzzling as cells use internal directional biases \cite{PNR13}, which increase persistence and migration speed, but may distract cells from sensing the direction of a gradient. This raises the question if cells employ some sort of memory   \textcolor{black}{ \cite{PNR14,PNR15}} as a form of active sensing strategy, in order to increase their sensing precision \textcolor{black}{(similar to  active membrane transport, which
improves transport, compared to passive transport).}
While memory is expected to improve precision in static environments, its benefit in changing natural environments is unclear.   
         
Cell sensing is performed through trans-membrane receptors which bind and unbind ligand molecules in the environment. At low ligand concentration, the precision is ultimately limited by the random arrival of ligand molecules on the cell surface by diffusion. An expression for this fundamental physical limit was first estimated by Berg and Purcell  \cite{PNR16} and was subsequently revisited  \cite{PNR17,PNR18, PNR18b}, considering a cell that tries to estimate the average occupancy of the receptor in a time interval  $\Delta t$ \textcolor{black}{(determined by slower downstream processes such as  cytoskeletal remodelling or rotary motor switching)}.  All these estimates include, in the single-receptor limit, the uncertainty due to ligand-receptor unbinding, which does not convey information about the ligand concentration. Instead, if a cell only registers binding events, it can further reduce the limit by a factor two \cite{PNR19}\textcolor{black}{, e.g. as potentially implemented by G-protein-coupled receptors with regulated signalling duration  \cite{PNR19a}, by endocytosis \cite{PNR19b}  or receptor diffusion \cite{PNR19c}}. 
A lower limit of the relative uncertainty (variance over mean) is estimated from the Cram{\' e}r-Rao bound of \textcolor{black}{ estimation} theory  
\BEQ
\label{PN1}
\frac{(\delta c)^2}{c^2}\ge \frac{1}{ c^2I(c)} \to \frac{1}{4Dac\Delta t}=\frac{1}{n},
\EEQ
where $I(c)$ is the Fisher information. In the limit of large  $\Delta t$ the right side of Eq.  \eqref{PN1} is obtained, with $c$ the  concentration, $D$ the diffusion coefficient, $a$ the receptor's linear dimension, and $n$ the number of binding/unbinding events within $\Delta t$. The result was then extended to time-dependent concentrations, i.e. ramp sensing  \textcolor{black}{ for a static cell or spatial gradient sensing for a moving cell  \cite{PNR20}}.  While more sophisticated than the above mentioned estimate of the signal-to-noise ratio in {\it Dictyostelium}  \cite{PNR11}, all these approaches neglect the receptor's history and hence memory. 

Experimental evidence for memory at the molecular level is widespread in biology,  \textcolor{black}{ and goes way beyond receptor methylation in bacterial chemotaxis  \cite{PNR21}. }During lactose uptake in {\it E. coli} the slow changes in the number of permease LacY lead to hysteresis \cite{PNR22}, evidence for a system's dependence on its past environment. Other examples include receptors during cell adhesion \cite{PNR23}, as well as neuronal plasticity and long-term potentiation, which can persist over days or even months  \cite{PNR24}. \textcolor{black}{Common to all these mechanisms is that they consume energy, e.g. in form of hydrolysis of adenosine triphosphate (ATP) or S-adenosyl methionine (SAM) \cite{PNR24b}.}

How does memory improve the precision of sensing? We start from considering the simple case of two consecutive measurements. The information from the first measurement is stored by the cell (e.g. via slow kinetic processes triggered by ligand binding  \cite{PNR23}) and is mathematically expressed by a prior distribution of the concentration values. Such prior information leads to a lower uncertainty in concentration sensing in the next measurement. Using the Bayesian Cram{\' e}r-Rao bound, the uncertainty is
\BEQ
\label{PN2}
(\delta c)^2\ge \frac{1}{ I(c)+I(\lambda)},
\EEQ
where $I(\lambda)$ is the Fisher information of  the prior contribution,   \textcolor{black}{with  $\lambda(c)$ a known prior distribution of the concentration $c$.  Applying} this inequality to a static environment,  \textcolor{black}{ we find that} memory of previous measurements allows an estimate with precision equivalent to a memory-less process with a correspondingly longer single-measurement time (see section  S2C of Supplementary Information (SI){{}}). The resulting precision therefore exceeds the physical limit for the single-measurement in Eq. \eqref{PN1}. Similar results have been obtained by others in static gradients \cite{PNR26,PNR27,PNR28}. However, cells in realistic environments may encounter concentrations which vary strongly in time e.g. by diffusion, flow, degradation by competing species, or variable chemical sources  \textcolor{black}{  \cite{PNR29,PNR30, PNR30b}}.

Does memory still help in fluctuating environments? As indefinite storage of past information and its analysis is impossible in cells, we anticipate an iterative sensing scheme at best. Our hypothesis is that the correlations of the environment should be important when considering memory since  \textcolor{black}{ long-range} correlations resemble static environments. In Fig. 1 we consider a fluctuating environment, with a changing concentration  \textcolor{black}{ slope (ramp or gradient)} at each time point, characterised by either correlated or uncorrelated fluctuations (depending on parameter $\alpha$). We propose two alternative schemes for sensing by a receptor using memory,  \textcolor{black}{known from control engineering}. In each scheme the receptor performs a measurement of both concentration  and  \textcolor{black}{ slope} in each time interval. This measurement is stored and the likelihood of subsequent concentration values is iteratively updated based on each new measurement. In one scheme, the receptor uses memory to predict the current environment, while in the other, the receptor carefully weighs past and current measurements to come up with an optimal estimate.  \textcolor{black}{ We finally 
provide preliminary evidence for 
 our predictions. 
Using  chemotaxis experiments on {\it Dictyostelium} amoeba in a microfluidic chamber and  spatio-temporal cell simulations, we find signatures of filtering in cell behaviour.}


\section*{ Models and Results}
{\bf Prediction and filtering schemes.} For sensing strategies with memory in fluctuating environments we use two iterative schemes: (i) The prediction scheme (Fig. 2a), in which the concentration is estimated {\em{ a priori}}, without the current measurement and only based on the history of previous measurements. This strategy might allow a cell to avoid toxins before encountering high harmful levels. (ii) The filtering scheme, in which the current concentration is optimally estimated based on previous and current measurements (Fig. 3a). This results in what is known as an {\em{a posteriori}} estimate. Such schemes are successfully adopted in control theory and in engineering applications (see \cite{PNR31} for an overview). For instance, the prediction scheme is used in weather and market forecast, as well as missile guidance. The filtering scheme is applied in navigation systems, where continuous update of the position based both on measurements and the system dynamics is of primary importance \textcolor{black}{(known as Kalman Filter)}. Famously, the navigation system of the landing lunar module of the Apollo 11 mission heavily relied on filtering-based software in the flight control system \cite{PNR32}. We first explain how the schemes work and derive analytical results, then consider how they might be implemented biochemically in cells. Finally, we discuss explicit biological examples, supporting our proposed schemes.

 Adopting a vectorial notation and indicating with ${\mathbf c}_t =(c^0, c^1)$ the concentration and  \textcolor{black}{ slope}, the concentration update at each time step can conveniently be expressed as (see Fig. 1 for further details)
 \BEQ
 \label{PN3}
 {\mathbf c}_t=F_t(\alpha){\mathbf c}_{t-\Delta t}+(1-\alpha){\mathbf w}_t
 \EEQ
 with 
\BEQ
\nonumber
 F_t= \left(\begin{array}{c c}
 1 & \Delta t\\
 0 & \alpha 
 \end{array} \right)
 \EEQ
 the matrix implementing the deterministic part of the evolution, $\alpha$ the  \textcolor{black}{ slope} correlation and ${\bf w}_t =(0, w_t)$ the fluctuation term with $w_t$  a random Gaussian variable with zero mean and variance $\sigma^2_w$. \textcolor{black}{ It is important to note that the concentration $c_t$ is unbounded. The advantage of this implementation is that for $\alpha=1$ the limit of a constant gradient is regained (slope fluctuations are  perfectly correlated at all times), while for $\alpha=0$ the slope fluctuations are completely uncorrelated. In contrast, any imposed limit on the concentration would induce correlations. }
 
  \textcolor{black}{In both schemes} the iteration step coincides with the single-measurement averaging time $\Delta t$. In this interval of time, measurement ${\bf y}_t =(y^0_t, y^1_t)$ with errors ${\bf \xi_t} =(\xi^0_t, \xi^1_t)$ is performed of both concentration  and \textcolor{black}{ slope, which are therefore assumed stationary in this interval.} We follow the protocol of \cite{PNR20} to connect the precision of the measurement to the number of binding events occurring in that interval (for concentration measurements, this is given by Eq. \eqref{PN1}). $\xi^0$ and $\xi^1$ are Gaussian random variables with zero mean and variances   $\sigma^2_{\xi^0}$ and $\sigma^2_{\xi^1}$ (see also Fig. 1 and section  S4B of SI  {{}}). 

\textcolor{black}{A key  parameter characterising the environment is $\alpha$, which correlates the slope} values at two consecutive times,  i.e. $\langle c^1_t \cdot c^1_{t+\Delta t} \rangle=\alpha \langle \left(c^1_t\right)^2\rangle$. As a result,  the autocorrelation of the  concentration $c^0$ is given by
\BEQ
\label{PN4}
\Phi(t+\tau,t)=
\underset{t\gg\tau\gg \textcolor{black}{\Delta t}}{\to}1-\frac{\tau}{2 t}+\frac{\alpha}{1-\alpha^2}\frac{\textcolor{black}{\Delta t}}{t},
\EEQ
valid in the limit of long time $t$ and large time separation 
\textcolor{black}{(note also that $\langle c^0_t\rangle=c^0_0$, for details see  Eqs. (S62-65) in SI).} Equation \eqref{PN4} shows that the environmental dynamics introduce a correlation in the concentration values that increases with $\alpha$. From Eq. \eqref{PN3} we can also evaluate the concentration variance at any time
\BEQ
\label{PN5}
\sigma^2_c (t)=\langle \left(c^0_t-c^0_0\right)^2\rangle \underset{t\gg \textcolor{black}{\Delta t}}{\to}\textcolor{black}{\left(t+\frac{2\alpha^2 \textcolor{black}{\Delta t} }{(1-\alpha)^2 (1+\alpha)}\right)\sigma^2_w},
\EEQ
which \textcolor{black}{increases} with increasing $\alpha$. Importantly, the concentration variance is asymptotically independent of  $\alpha$, thus allowing us to compare fluctuating environments with different correlation but same variance.

Figure 2a summarises the steps leading to the iterative update of the covariance matrix with memory in the prediction scheme. 
\textcolor{black}{
At any given time one can define the {\it a priori}   \textcolor{black}{$\tilde{P}_t$ }and { \it a posteriori}   \textcolor{black}{$\hat{P}_t$} covariance matrices as follows:}
\textcolor{black}{\begin{subequations}
\begin{align}
\tilde{P}_t &=\langle \delta \tilde{\bold{c}}_t  \delta \tilde{\bold{c}}_t^T \rangle \label{covsa}\\ 
\hat{P}_t  &= \langle \delta \hat{\bold{c}}_t  \delta \hat{\bold{c}}_t^T \rangle \label{covsb},
\end{align}
\end{subequations}
where  $\tilde{\bold{c}}_t$  and   $ \hat{\bold{c}}_t$  indicate the estimates of the concentration and slope performed before  \textcolor{black}{({\it a priori})} and after  \textcolor{black}{({\it a posteriori})} making a measurement, respectively, and $\delta\tilde{\bold{c}}_t$ and  $\delta \hat{\bold{c}}_t$  are the differences of these estimates from the true values  ${\bold{c}}_t$.
From Eq. (3) it follows:
\textcolor{black}{
\BEQ
{\tilde{\mathbf c}}_t=F_t(\alpha){\hat{\mathbf c}}_{t-\Delta t} 
\EEQ }
and therefore, by Eq. (3) and the definitions Eqs. (\ref{covsa}) and (\ref{covsb})
\BEQ
\label{recur1}
\tilde{P}_t=F_t \hat{P}_{t-\textcolor{black}{\Delta t}} F_{t}^T +Q_t,
\EEQ
where  $Q_t$ is a  diagonal matrix carrying the variance $(1-\alpha)^2 \sigma_w^2$ of the slope fluctuations (see also Eq. (S61e) in SI).
Due to the Cram{\' e}r-Rao bound,
\BEQ
\label{CRm}
\hat{P}_{t\textcolor{black}{-\Delta t}} \geq \left(\tilde{P}_{t-\Delta t}^{-1}+R_{t-\Delta t}^{-1}\right)^{-1},
\EEQ
where  $R_t$ is a  diagonal matrix carrying the variances $\sigma^2_{\xi^0}$ and $\sigma^2_{\xi^1}$ of the single-measurement errors on concentration and slope (see  for details section S4 and  Eq. (S61b) in SI).
Replacing Eq. (\ref{CRm}) into Eq. (\ref{recur1}) leads to the following recursive relation for the  {\it a priori} covariance matrix:
\BEQ
\label{recurP}
\tilde{P}_t =F_t \left(\tilde{P}_{t-\Delta t}^{-1}+R_{t-\Delta t}^{-1}\right)^{-1} F_t^T+Q_t,
\EEQ
which can be solved analytically at steady state.}

The filtering scheme is summarised in Fig. 3a. \textcolor{black}{ In this scheme an {\it a posteriori} estimate of the concentration  $ \hat{{\bold c}}_t$   is obtained
as a linear combination of the  {\it a priori} estimate $\tilde{{\mathbf c}}_{t}$ and the  direct measurement ${\mathbf y}_t$
\BEQ
\label{coveq}
 \hat{{\mathbf c}}_t=\tilde{{\mathbf c}}_{t}+ M_t \left({\mathbf y}_t-\tilde{{\mathbf c}}_{t} \right),
 \EEQ
in which the weight $M_t$ is chosen so as to minimise the covariance {\it a posteriori} matrix (see Eq. (S93) in SI).
With such choice for $M_t$ from Eq. (\ref{coveq})  it  follows
 \BEQ
\label{Kal1}
\hat{P}_t=(\mathcal{I}-M_t)\tilde{P}_t.
\EEQ
Inserting  Eq.~(\ref{recur1}) into (\ref{Kal1}) leads to the following recursive relation for the {\it a posteriori} covariance matrix in the filtering scheme
 \BEQ
 \label{Kalm}
 \hat{P}_t=(\mathcal{I}-M_t)\left(F_t \hat{P}_{t-\Delta t} F_{t}^T +Q_t \right).
 \EEQ
}
 We obtained analytical expressions for the stationary solution of the covariance matrices by imposing convergence of the iterative relations. From these solutions we extracted the uncertainties for the concentration and  \textcolor{black}{ slope }in both schemes (see also section s S4C and S4E of SI). \textcolor{black}{ Importantly, while} the dynamics of $c^0_t$ is diffusive and therefore non-stationary (see Eqs. \eqref{PN4} and \eqref{PN5}), the dynamics of the increments $c^1_t$ is stationary with correlations decaying as $\alpha^{\tau}$ with $\tau$  the time difference. This stationarity in the concentration increments ultimately allows the convergence of the iterative schemes.  Figures 2b and 3b demonstrate simple biochemical implementations of the two schemes to be discussed later.

Figure 4 shows plots of the total uncertainties   \textcolor{black}{ $\Delta_T^P$ for the prediction (a) and $\Delta_T^F$  for the filtering scheme (b), both defined as the trace of the respective steady-state covariance matrices (for individual uncertainties of concentration and   slope sensing see Fig. S3 in SI). Both uncertainties are expressed in units of the total single-measurement error defined as:
\BEQ
\Delta_T^S =\sigma^2_{\xi^0} +\sigma^2_{\xi^1} \Delta t^2
\EEQ
with $\sigma_{\xi^0}$ and $\sigma_{\xi^1} $  the errors on concentration and slope, respectively, for a single measurement performed in time $\Delta t$.}

 Both the effects of correlation $\alpha$ and the magnitude of environment-to-noise ratio (ENR) $\eta=\sigma_w/\sigma_{\xi^0}$ are shown. The latter is defined as the ratio between the to-be-measured amplitude of the  \textcolor{black}{ slope} fluctuations and the single-measurement error without memory. The main noticeable difference between the two schemes is that in the filtering scheme the total uncertainty is always lower than the single-measurement error, while in the prediction scheme the total uncertainty exceeds the corresponding single-measurement error for large ENR, i.e. when the environment fluctuates strongly compared to the single-measurement error.  \textcolor{black}{ Hence,  while predicting an unpredictable environment is impossible, when filtering strongly fluctuating environments one can always rely on single-measurements. }


 \textcolor{black}{ Note that while we set the times of the slope fluctuations and the measurement equal in our derivation, our results can be generalised. In fact, slower fluctuations would allow more measurements to be conducted  between changes in the slope, which due to Eq. \eqref{PN2} can be mapped onto a single-measurement with a longer rescaled measurement time.  Since the solutions for the uncertainties scale with the single-measurement error  (see also Eq. \eqref{PN6} below), this leads to an overall reduction, proportional to the ratio of fluctuations to measurement time.
   However,  when normalised to the total single-measurement error,  the plots in Fig. 4 remain unchanged. }

\noindent {\bf Analytical results for uncertainties.}
A simple illustrative expression of the uncertainties in Figs. 4a and 4b is given by the solution for the uncorrelated case ($\alpha=0$) for which the covariance matrix is diagonal in both schemes (for general case see section  S4 and Fig. S3 in SI). For the filtering scheme we find that the total uncertainty is
\BEQ
\label{PN6}
\Delta^F_T=\frac{\kappa^2}{2\left(1+\frac{\kappa^2}{\eta^2}\right)}\left[1+\sqrt{1+\frac{4}{\kappa^2}\left(1+\frac{\kappa^2}{\eta^2}\right)}\right]\sigma^2_{\xi^0},
\EEQ
while for the prediction scheme we obtain $\Delta^P_T=\Delta^F_T+\sigma^2_w$. Without  \textcolor{black}{ the} current measurement, the uncertainty due to the random slope cannot be eliminated and adds up to the uncertainty in the filtering scheme. $\kappa$ is the ratio between the measurement errors on $c^1$  and $c^0$, i.e. $\kappa=\sigma_{\xi^1}/\sigma_{\xi^0}$, in units of  $\Delta t$.

Equation \eqref{PN6} is best understood in certain limits. For very accurate measurements of the \textcolor{black}{ slope}  ($\kappa \to 0$), filtering returns a perfectly accurate estimate for the concentration, i.e. with zero uncertainty due to optimal predictions based on the exactly measured  \textcolor{black}{ slope}. The opposite limit ($\kappa \to \infty$) corresponds to a larger uncertainty, as expected, but always smaller than the single-measurement error. The latter is achieved as an upper limit only in the further limit of large environmental fluctuations, i.e. $\sigma_w \to \infty$. This is because, for very large fluctuations of the environment and very inaccurate  \textcolor{black}{ slope} measurements, no more information is collected from past measurements than is obtained in a single-measurement (see also Figs. 4a and 4b). The same behaviour is shown also for correlated environments ($\alpha>0$), with the general feature that in this case the iterative filtering scheme always leads to a decreasing total uncertainty for increasing value of the correlation parameter $\alpha$ (Fig. 4b).  Therefore, memory always allows the receptor precision to go beyond the physical limit of the single-measurement, with larger improvement for environments with more correlated fluctuations.  

When is it better to predict based on memory and when is it better to brute-force measure? For the prediction scheme, the cross-over regime can be estimated from the value at which the \textcolor{black}{ slope}  uncertainty equals the single-measurement error (see Fig. 4a and section  S4C in SI). This value is given by $(1-\alpha)^2\sigma_w^2=\kappa^2 \sigma^2_{\xi^0}$, which is exact for an uncorrelated environment ($\alpha=0$ and $\kappa \gg1$) and is approximately valid for correlated environments (Fig. S4).  This cross-over simply reflects the fact that without the current measurement any prediction will retain the whole uncertainty  due to  \textcolor{black}{slope} fluctuations. When this is of the same order of the single-measurement error in the  \textcolor{black}{ slope}, the final estimate based on past measurements will be less accurate than the total single-measurement. While inferior to the filtering scheme, the prediction scheme could be beneficial when trying to avoid certain environments.

\noindent {\bf Biochemical implementation in cellular networks.}
How can the two schemes be implemented biochemically in cell-signalling networks? \textcolor{black}{We propose two simple pathways in the continuous limit, which show qualitatively similar behaviour to the two discrete schemes from above. } In Fig. 2b  simple production of a species $y$ upon activation of the receptor  \textcolor{black}{ according to 
\BEQ
\label{bio1}
\dot{y}_t=k_y(u_t-y_t)
\EEQ
 allows the cell to monitor the input with a ``delay''.}
This produces the prediction scheme, in which the current-time value of the concentration is not accessible.  For slowly varying inputs in $c$, $ y $ can follow $c$ while averaging out measurement noise of the activity $u$. In Fig. 3b an incoherent feedforward loop implements the filtering scheme. In this case, second species $x$, produced upon activation of the receptor, inhibits $y$ \textcolor{black}{according to 
\begin{align}
\label{bio2}
\dot{x}_t &=k_x(u_t-x_t)\\
\nn \dot{y}_t &=k_y \left(\frac{u_t}{x_t}-y_t \right).
\end{align}
}
 In addition to filtering out measurement noise the combined monitoring of the two species allows the cell to follow the input, even when it rapidly changes. 

Figures 4c and 4d show the results of the simple biochemical prediction and filtering schemes from Figs. 2b and 3b, respectively. Here, the input is the ligand concentration, whose dynamics are described by a stochastic Poisson process with correlation time $\lambda_s^{-1}$ and amplitude $2\lambda_s$ (for details see section  S5 in SI). Plots share the same properties as the schemes from Figs. 4a and 4b with the uncertainty decreasing with increasing correlation time/decreasing amplitude. Specifically, Fig. 4c shows that molecule $y$ can follow slowly/weakly fluctuating inputs with a reduction in uncertainty relative to the single-measurement noise due to time averaging by the slow kinetics of $y$. For rapid/strongly fluctuating inputs $y$ cannot follow anymore, is out of phase and hence becomes worse than the single-measurement error. In contrast, Fig. 4d shows that the incoherent feedforward loop of the two molecular species, $x$ and $y$, always leads to a reduction in uncertainty (depending on suitable parameter choices, see section  S5 in SI for details). If the input fluctuates slowly/weakly, $y$ adapts adiabatically and $x$ tracks the input accurately, with the measurement error in $u$ effectively filtered out. If the input fluctuates fast/strongly, $x$ cannot follow the input anymore. However, $y$ with faster dynamics than $x$ can follow the input now and filter out noise. Hence, this network always leads to a reduction in the measurement error.
\vspace{0.2cm}
 
 \noindent{\bf Experimental 
 support for filtering in  \textcolor{black}{  {\textbf {\emph {\normalsize Dictyostelium} }}chemotaxis.}}
 As filtering is the more advanced scheme of the two, is there any evidence  of cells using this strategy? \textcolor{black}{ In filtering the weight  $M_t$ of the current measurement in the final estimate of the concentration and slope is a key feature (cf. Eqs. (\ref{coveq}-\ref{Kalm})).
  In particular, the weight of the current measurement for the slope, defined as the second diagonal element of the matrix $M_t$ at stationarity given by $\hat{P}^{11}/(\kappa \sigma_{\xi^0})^2$ (see Eq. (S115)), is close to $1$ for large fluctuations in the environment and close to zero for small fluctuations. In the latter case, more accurate predictions can be made based on memory of previous values of concentration and slope.} 
  
  Figure 5a shows the weight of the current  \textcolor{black}{ slope }measurement in the filtering scheme (for details see section  S6 in SI). For small  \textcolor{black}{environmental} fluctuations the weight of current measurement is small, since in this regime the cell can rely on memory to improve the concentration estimation, while for large environment fluctuations the prediction based on memory of previous measurements becomes unreliable and the weight shifts toward the current measurement. The weight of the current measurement contributes therefore significantly for large fluctuations with little correlation (large ENR and small $\alpha$, respectively).

 Testing these predictions of filtering in cell sensing  \textcolor{black}{ is difficult for individual receptors, in particular since chemotaxis is an emergent phenomenon arising from a large number of  biophysical and biochemical details  \cite{PNR33}. Instead,} we approached this problem by conducting chemotaxis experiments on starved {\it Dictyostelium discoideum} cells in a microfluidic chamber  \textcolor{black}{ as described in  \cite{PNR34}}. In this setting  \textcolor{black}{ with a left and right  inflow,} it is possible to study the turning behaviour of the migrating cells when the gradient in cAMP concentration is instantaneously switched in direction \textcolor{black}{  from left to right and vice versa}.  Specifically, we use the recovery of the chemotactic index ($\Delta CI$), with $CI$ a measure of the alignment of the cell movement with the gradient, as an indicator for turning speed and a cell's reliance on its current measurement (see Materials and Methods for details). Considering multiple cells in different concentration gradients, Fig. 5b \textcolor{black}{ (solid line) }shows that this recovery is directly correlated with gradient steepness (i.e. amplitude of fluctuations or ENR). Exemplar turning behaviours are shown in Figs. 5c for slow  \textcolor{black}{ (left)} and fast  \textcolor{black}{ (right) }turning cells, respectively. In line with our prediction from filtering, the fraction of fast turning cells indeed increases with ENR, indicative of cells trusting their current measurement in strongly fluctuating environments (Fig. 5e\textcolor{black}{, left}). \textcolor{black}{While this observation is not sole proof of filtering, numerous other observations point in the same direction (see Discussion section)}.
 \\
\noindent  \textcolor{black}{ {\bf Spatio-temporal simulations explain observed cell behaviour.} Although the filtering observed in the turning behaviour of {\it Dictyostelium} cells cannot be attributed to a single pathway or a small number of molecular species \cite{PNR14,PNR15,PNR33}, it is worth pointing out that the activity  of RasG downstream of the cAR1 G-protein-coupled receptor follows an incoherent feedforward loop and helps mediate adaptation to persistent cAMP stimulation  \cite{PNR35}. Hence, to understand if a spatially extended incoherent feedforward loop can reproduce our data, we implemented a minimal biophysical-biochemical model of pseudopod-guided chemotaxis. }
  \noindent \textcolor{black}{ Following  \cite{PNR33,PNR33b} we coupled a 2-dimensional membrane with a Meinhardt-like reaction-diffusion system that exerts driving forces on the membrane biased by external cues. This was combined with an incoherent feedforward loop to process sensory inputs  \cite{PNR36}. In analogy with the experiments on {\it Dictyostelium}, we then observed the turning behaviour of these simulations in response to a reversal in gradient direction with different ENRs. Similarly to our live-cell experiments, simulated cells quickly reoriented themselves in response to high-ENR \textcolor{black}{switches}, with reorientation time increasing as environmental fluctuations became smaller (Fig. 5b, dotted line). Cell turning circles in response to small-\textcolor{black}{ENR} switches (Fig. 5d, left) were considerably larger than in response to large-\textcolor{black}{ENR} switches (Fig. 5d, right), mirroring the behaviour of live cells (Fig. 5c). Similarly, the fraction of fast-turning cells (Fig. 5e, right) matches the data (Fig. 5e, left).
 }
 \section*{Discussion}
 We presented an analytical calculation for two active sensing strategies with memory in fluctuating environments, termed prediction and filtering. Importantly, in correlated environments both strategies would allow cells to sense far more precisely than predicted by current estimates of the fundamental physical limit, thus providing potential explanations for the observed single-molecule precision in chemo-sensing {\it E. coli}, neurons, and T cells. Filtering, the more elaborate of the two schemes, can improve the precision of sensing even in uncorrelated, difficult-to-predict environments.  Through  direct observation of the turning behaviour of chemotacting {\it Dictyostelium} cells  in response to fluctuating  cAMP gradients in microfluidic chamber \textcolor{black}{(and through previous evidence provided below)}, we found support for the adoption of filtering by these cells.
 Hence, cells do not simply extend a pseudopod in the estimated gradient direction based on their current measurements \textcolor{black}{(traditional compass model)}. 
 Instead, cells   weigh  past and current measurements in chemotaxis, resulting in an adjustment of the turning radius. \textcolor{black}{ This matches our intuition that a smaller change in the environment requires a longer time for the cell to notice.}
  Biochemically, filtering can be implemented by the incoherent feedforward loop,  \textcolor{black}{ both at the single-receptor (Figs. 3b and 4c) and whole-cell (Fig. 5d) level. }Indeed, recent experiments demonstrate that such a feedback loop is implemented by the small GTPase Ras, which may be responsible for adaptation in {\it Dictyostelium} cells  \cite{PNR35}.

  \textcolor{black}{  While our prediction and filtering schemes have a long history in control engineering \cite{PNR31,PNR32, PNR40b}, their application to fluctuating gradient sensing constitutes a new direction. Recently, prediction was investigated in anticipating oscillations, e.g. as part of the circadian clock, requiring energy dissipation \cite{PNR40c} in line with our active sensing strategies. Note ``active sensing'' refers to the need for energy consumption (similar to active transport in cells) and must not be confused with ``active learning'' in artificial intelligence and machine learning. Such active processes can also be used to time-average noisy signals to enhance the accuracy in sensing \cite{PNR40cc}. The Kalman filter was previously applied to adaptation in bacterial chemotaxis  \cite{PNR40d}, demonstrating  close connection to integral feedback control. This filter was however not  considered in the context of the physical limits of sensing. The Kalman filter can be considered the most simple dynamic Bayesian network \cite{PNR40e}. Note the cell's parameters, e.g. rate constants for optimal weighting of past and current measurements, would most likely need to be adjusted by evolution to match typically encountered stimuli.}

 Our results may provide new insight into previously observed internal directional biases in immobilised latrunculin-treated cells, dynamically   stimulated by uncaging cAMP using a circular UV beam  \cite{PNR13}.  The precision of directional cell sensing was found to be determined by a combination of external cAMP sensing and internal bias of unidentified origin. Since internal biases not aligned with the external gradient are expected to reduce the sensing precision, as confirmed by theoretical modelling  \cite{PNR27}, the purpose of 
 internal biases remained unclear. While a beneficial increase in cell heterogeneity is a possibility (at the expense of precision) \cite{PNR13}, our work suggests that in a dynamic environment the internal bias represents memory for filtering, leading to an actual increase in sensing precision.

 Observations of memory and filtering go back at least four decades  \cite{PNR37,PNR38,PNR39}. In \cite{PNR38} neutrophils filtered out temporally-changing concentrations of N-formyl-methionyl-L-leucyl-phenylalanine (fMLP) over a $10s$ time scale. Although no connection with sensing precision was made, two cell fractions of turning behaviour were observed -- a slowly U-turning and a fast $180^{\circ}$-repolarising fraction. Most recently  \cite{PNR39} {\it Dictyostelium} cells responded to \textcolor{black}{ramps}  (their Fig. 4e), as expected from the incoherent feedforward loop
and closely related networks  \cite{PNR20}. Consistently, pulses of high frequency (14 mHz) were filtered out but not pulses of low frequency (7 mHz) (their Fig. 4a). Even {\it E. coli} bacteria were observed to filter in response   to alternating gradients, producing coherent waves at the population level when stimulated at $0.01$ Hz \cite{PNR40}. Hence, our results integrate a number of apparently distinct observations across different cell types.
 
Wolpert's  ``no free lunch theorem'',  \textcolor{black}{ used as a metaphor here,} states that to optimally estimate, search, etc., prior assumptions are necessary  \cite{PNR41}. \textcolor{black}{In effect, there is no ``short cut'' for a solution and to do better on average a ``cost'' needs to be paid.  Indeed, our two active sensing strategies  are based on such priors (memory), updated dynamically and the presence of structure in the environment (correlations) allows one to outperform a simple direct counting algorithm (i.e. single measurement)}. Active sensing strategies are widespread in biology, including integral feedback control in bacterial chemotaxis  \cite{PNR42}, olfactory- and photo-transduction  \cite{PNR43}, as well as kinetic proofreading  \cite{PNR44}. The latter is a form of  ``error correction'' mechanism 
for enhancing the specificity in DNA replication, protein synthesis, homologous recombination  \cite{PNR45} and T-cell signalling  \cite{PNR46}.  The striking similarity between our proposed strategies, macroscopic engineering solutions \cite{PNR32}, and learning by Bayesian inference in humans  \cite{PNR47} hint toward universal problem solving strategies in nature.
\section*{Methods}

  

\noindent  {\bf Chemotaxis experiments and image analysis.} The migration of  \textcolor{black}{ $93$} starved AX2 strain {\em Dictyostelium discoideum} cells was recorded  \textcolor{black}{ as described in  \cite{PNR33,PNR34}. Imaging was performed } by differential interference contrast (DIC) in a $\mu$-slide 3-in-1 microfluidic chamber (Ibidi) with three $0.4\times1.0$mm$^2$ inlets that converge at an angle of $32^{\circ}$ to the main channel (dimension $0.4 \times 3.0 \times 23.7 $mm$^3)$. Two micrometer valves (Upchurch Scientific) reduced the flow velocities from the side inlets. The central inlet was connected to an infusion syringe pump (TSE Systems), which generated a stable flow of 1ml/h. Inlets contained a mixture of the chemoattractant cyclic adenosine monophosphate (cAMP) and an Alexa Fluor red dye of similar molecular weight (Invitrogen), to allow the characterisation of gradients (for details, see \cite{PNR34}).  \textcolor{black}{ Microscopy was} performed using an Axiovert 135 TV microscope (Zeiss), with LD Plan-Neofluar objectives 20x/0.50NA and 40x/0.75NA (Zeiss) in combination with a DV2 DualView system (Photometrics). Images were taken at a frame rate of $1/3$ sec.  \textcolor{black}{ Cells were given time to adjust to stable chemical gradients for about 20 min before the direction of the gradient was switched. Typical trajectory 
 lengths were 37 min.} Cell outlines and centres of mass were extracted using a custom-written plug-in for ImageJ. Chemotactic indices of cells are given by $CI = \cos(q_m - q_g),$ where $q_m$ is the angle of migration and $q_g$ is the angle of the gradient. $CI$s were measured over a period of 2 min on either side of each switch, with the change in $CI$ being defined as $\Delta CI = CI_{\text{after}} -CI_{\text{before}}$. 

 \noindent  {\bf Cell simulations.}   \textcolor{black}{Sensory information is processed by an incoherent feedforward loop \cite{PNR35}, consisting of a response component and its inhibitor, both driven by the outside stimulus. The response component is passed to a Meinhardt model simulated on a 2-dimensional deformable membrane \cite{PNR33, PNR33b}. The Meinhardt model consists of a self-promoting local activator that inhibits activation elsewhere on the membrane via a global inhibitor. Patches of activator also self-limit by driving the production of a local inhibitor, allowing new patches to form nearby. The membrane moves through the combination of an outward normal force proportional to the local activator concentration, global membrane tension, local bending tension and a normal, area conserving force representing cytosolic pressure. See  section  S7 in SI {{}} for details including differential equations and parameter values.}

\begin{acknowledgments}
{\small
We thank Peter Swain and Martin Howard for insightful feedback and B$\ddot{\mathrm{o}}$rn Meier for contributing data. We thankfully acknowledge support by the Leverhulme-Trust, Grant N. RPG-181 and European Research Council Starting-Grant N. 280492-PPHPI.}
\end{acknowledgments}
\section*{Author Contributions}
\small{
 G. A., L.T., and R.G.E. conceived and designed the project. G. A. developed the theory and numerical 
algorithms. L. T.  conducted simulations and D. H. provided the data. All authors analysed the results and wrote the manuscript.
\section*{Additional Information}
\small{
\noindent {\bf Supplementary Information} accompanies this paper at http://www.nature.com/
scientificreports\\
{\bf Competing financial interests:} The authors declare no competing  financial interests.\\
{\bf How to cite this article}: }




\newpage
\begin{figure*}[h]
\centerline{\includegraphics[width=0.9\linewidth]{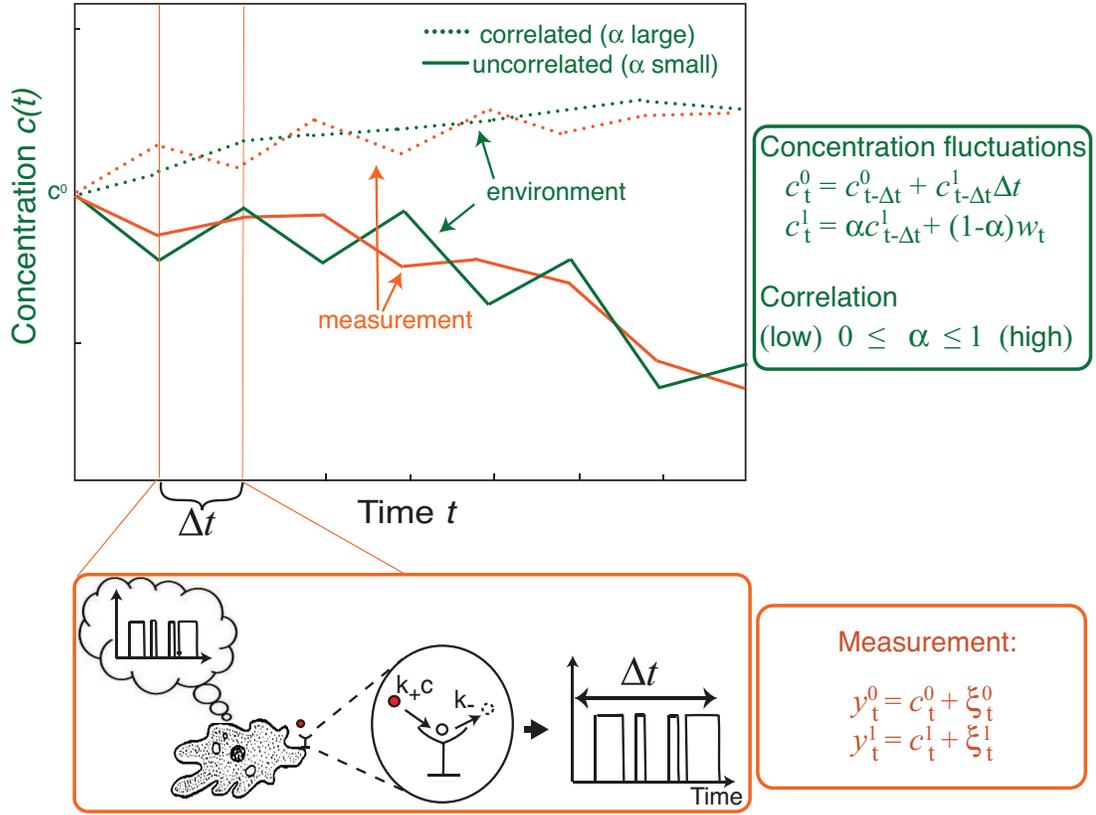}}
 \singlespace\caption{{\bf Fluctuating environments of chemical concentration and single-receptor measurement.} (top left) Exemplar concentration profiles for two different values of the \textcolor{black}{slope}  correlation, $\alpha=0.1$ (green continuous lines) and $\alpha=0.9$ (green dotted lines), corresponding to small and high correlations, respectively. (Orange continuous and dotted lines) Corresponding sequences of measurements, performed by the cell in each interval of time $\Delta t$, for the two concentration profiles. (top right, green equation box) Iterative dynamics generating the concentration profiles. At each time step the concentration  $c^0$ and  \textcolor{black}{slope} $c^1$ are updated. Parameter $\alpha$ ($0<\alpha<1$) represents the correlation strength of the \textcolor{black}{slope}  fluctuations $w_t$, a Gaussian variable with zero mean and variance $\sigma^2_w$. For $\alpha=0$ the profile is a fluctuating slope with totally uncorrelated fluctuations. For $\alpha=1$ the fluctuations vanish and the profile becomes a constant slope. (bottom left, orange box)  Illustration of the single-measurement performed by a cell-surface receptor in the time interval $\Delta t$. The receptor records ligand binding and unbinding  events occurring with respectively rates $k_+ c(t)$ and $k_-$, and estimates the average concentration from the time series of receptor occupancy  (cloud in illustration represents memory). (bottom right, orange equation box) $y^0$ and $y^1$ are estimates of concentration  and slope in time interval $\Delta t$. $\xi^0$ and $\xi^1$ are stochastic Gaussian variables with zero mean and variances $\sigma^2_{ \xi^0}$  and  $\sigma^2_{\xi^1}$  given respectively by $(c^0)^2/n$ and $12(c^0/\Delta t)^2/n$ \cite{PNR20}, the limits for the precision of concentration and \textcolor{black}{ slope}  sensing  in time interval $\Delta t$ with $n$  binding/unbinding events  (Eq. (S37) in SI). }\label{Fig1}
\end{figure*}

\begin{figure*}[h]
\hspace{-6cm}
\centerline{\includegraphics[width=0.9\linewidth]{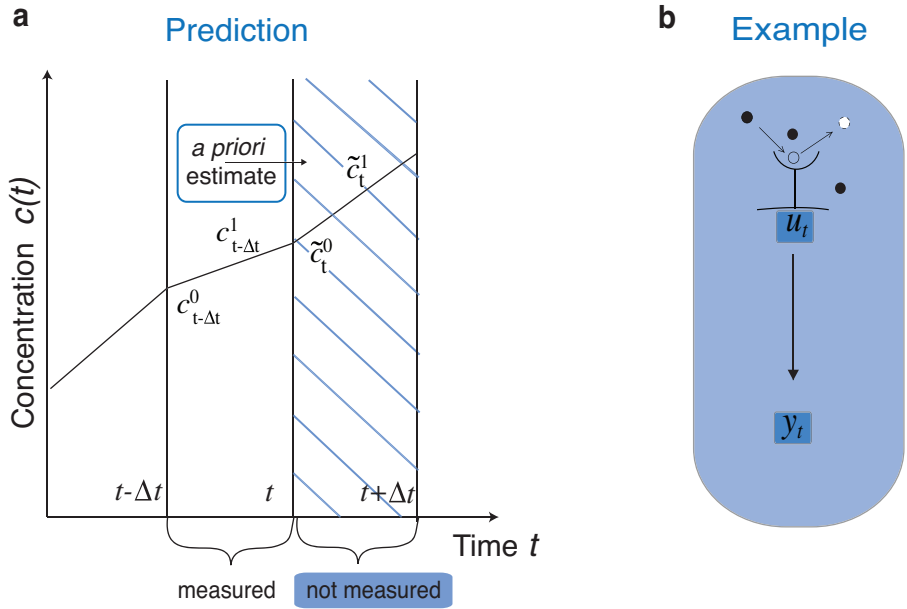}}
 \singlespace\caption{{\bf Iterative prediction scheme for receptor sensing with memory.} (a) At each time step the concentration profile is measured and updated  (see also green and orange equation boxes in Fig. 1). Receptor sensing is based only on past measurements to make an {\it a priori} estimate (prediction) for both  concentration $\tilde{c}^0$    and  slope $\tilde{c}^1$  with accuracy  limited by the Cram{\' e}r-Rao bound.
  (b) Simplest implementation of the scheme in a cell-signalling network. Upon binding and unbinding of external ligand, the receptor activity $u(t)$ induces production of molecule species $y$.  For slow reaction rate, the production of $y$ mirrors the input with a delay, reproducing the condition of the prediction scheme in which the current time measurement is not accessible. For details see sections S4A and S5A in SI. }\label{Fig2}
\end{figure*}

\begin{figure*}[h]
\centerline{\includegraphics[width=0.9\linewidth]{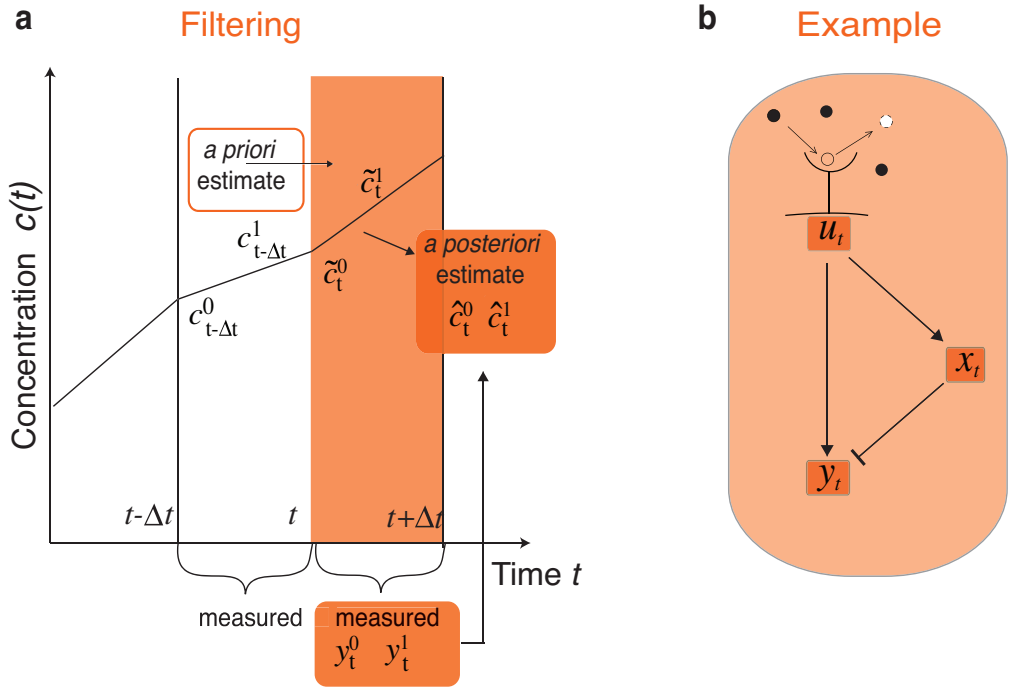}}
 \singlespace \caption{{\bf Iterative filtering scheme for receptor sensing with memory.} (a) The {\em a priori} estimate ${\bf \tilde{c}}_t=(\tilde{c}^0,\tilde{c}^1)$   is corrected by including the current time measurement ${\bf y_t}$ with a weight $M_t$   that minimises the {\em a posteriori} covariance matrix $\hat{P}_t$.   
(b) Simplest implementation of the scheme in a cell-signalling network. In the incoherent feedforward loop species $y$ is inhibited by $x$. The combined monitoring by the two species allows accurate tracking of $c$ while filtering measurement error in $u$. For details see section s S4D and S5B in SI.}\label{Fig2}
\end{figure*}

\begin{figure*}[h]
\vspace{-1cm}
\centerline{\includegraphics[width=0.85\linewidth]{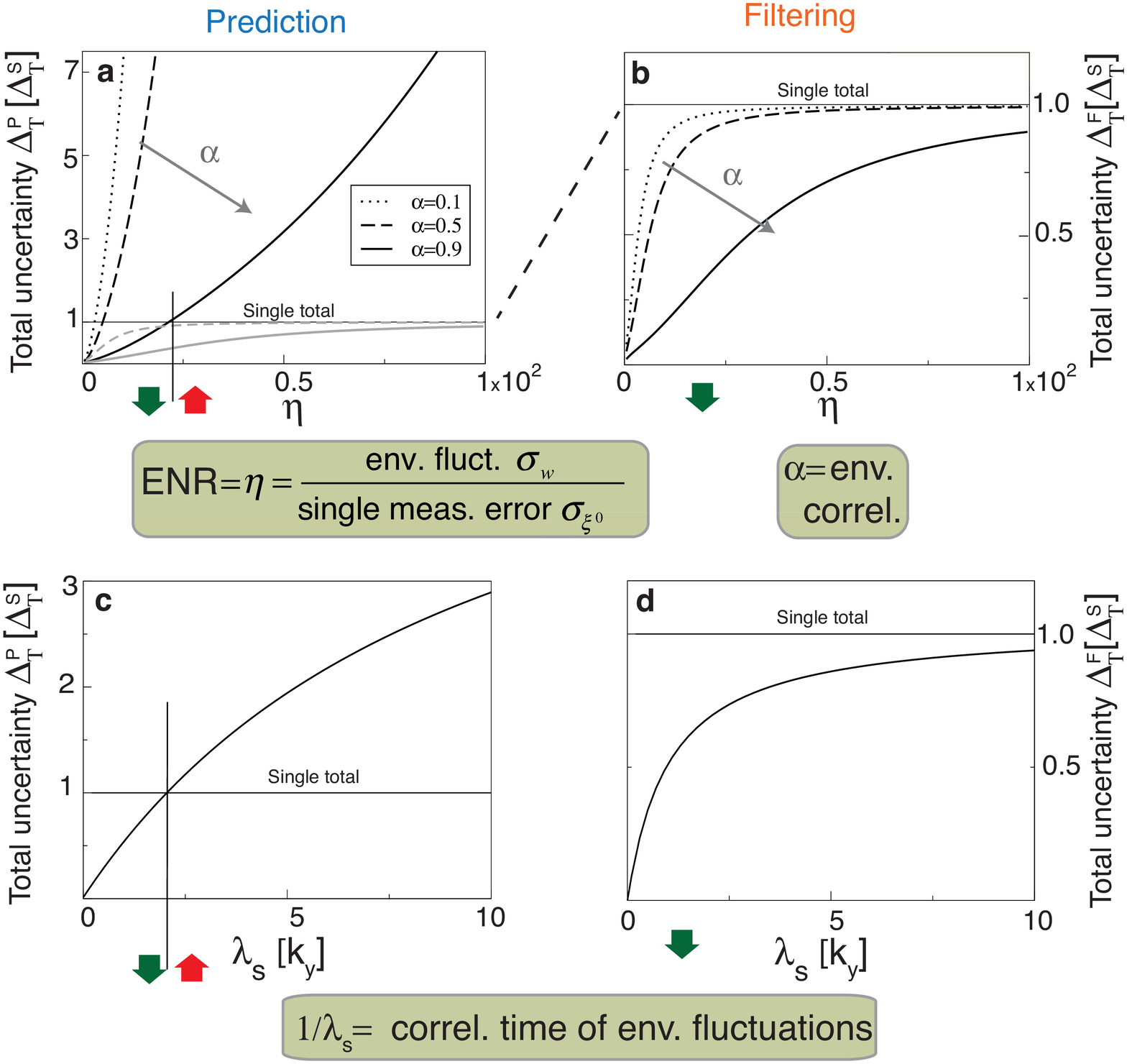}}
\vspace{-0.6 cm}
 \singlespace \caption{{\bf Performance of prediction and filtering schemes.}  (a) Prediction scheme. Total uncertainty $\Delta^P_T$ \textcolor{black}{(defined in the text)} as obtained from the trace of the stationary solution for the {\em a priori} covariance matrix plotted as a function of the ENR $\eta=\sigma_w/\sigma_{\xi^0}$,   i.e. the ratio between   \textcolor{black}{slope} fluctuations with width $\sigma_w$   and  concentration single-measurement error $\sigma_{\xi^0}$, for three values of the environment correlation $\alpha=0.1, 0.5$ and $0.9$ (grey curves  are introduced for comparison with the filtering  scheme, see (b)). Without a current estimate for the  \textcolor{black}{slope}, a prediction cannot be more precise than $\sigma_w$   and with increasing $\eta$ the uncertainty necessarily crosses over from a regime which it is smaller than the single-measurement error (green arrow) to a regime in which it is larger (red arrow). (b) Filtering scheme. Analogous plots of $\Delta^F_T$  \textcolor{black}{ given by} the trace of the {\em a posteriori} covariance matrix. While the uncertainties increase with ENR, they remain always lower than the respective single-measurement errors (green arrow) due to access to the current measurement of the  \textcolor{black}{slope}   (see text and section  S4 in SI for details). Parameters: following \cite{PNR20} we take $\sigma_{\xi^1}=\kappa \sigma_{\xi^0}$ with $\kappa=\sqrt{12}$. Ligand concentration $c^0=0.01$, $\sigma_{\xi^0}=(c^0)^2$, and $\Delta t=1$ (arbitrary units). Uncertainties are in units of the total single-measurement error $\Delta^S_T=\sigma^2_{\xi^0}+\sigma^2_{\xi^1}\Delta t^2=(1+\kappa^2)\sigma^2_{\xi^0}$. (c-d) Implementation of the prediction (c) and filtering (d) schemes with the two cell-signalling modules of Figs. 2b and 3b. Total measurement uncertainty $\Delta^{P/F}_T$,  \textcolor{black}{defined in section S5 in  SI  as a function of} the difference between the signal transmitted through the modules including $(\delta \sigma)$ and excluding $(\delta \sigma')$  noise, is plotted as a function of the inverse correlation time  (equivalent to the amplitude in the input activity $\delta u$). The module parameters for  $\delta \sigma$ and $\delta \sigma'$ are such that input noise is filtered out (slow rates) and the output follows exactly the input (fast rates), respectively. This difference  (averaged over noise) measures the accuracy of the two modules in transmitting the input signal, and reproduces the prediction and filtering schemes in (a) and (b), respectively. Uncertainties are in units of   $\Delta^{S}_T$, 
   \textcolor{black}{defined in section S5 in SI as a function of the difference between 
   $\delta \sigma''$  and  $\delta \sigma$},  with $\delta \sigma''$ equal to  $\delta \sigma$  but evaluated with the same module parameters (fast rates) used for $\delta \sigma'$ (no noise filtering). In (c) $\delta \sigma=\delta y$ and in (d)  $\delta \sigma=\delta x+\delta y$  with $\delta x$ and $\delta y$ deviations from steady-states values (see section  S5 in SI for details).  The horizontal thin line in each plot indicates the total uncertainty of the single-measurement. }\label{Fig4}
\end{figure*}

\pagenumbering{gobble}
\begin{figure*}[h]
\vspace{-1 cm}
\hspace{-2.7 cm}
\centerline{\includegraphics[width=0.6\linewidth]{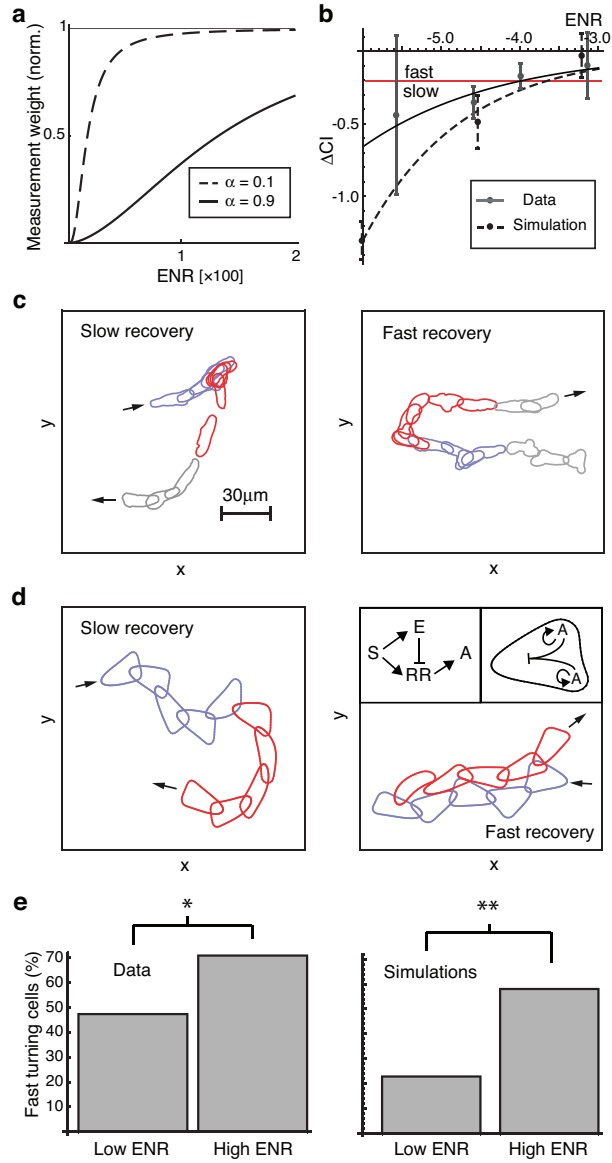}}
\vspace{-0.0cm}
  \singlespace \caption{{\bf Amoebae may use filtering in fluctuating environments. }
  (a) The weight of the cell to favour the current  \textcolor{black}{slope}   measurement over their prediction based on memory is plotted for $\alpha=0.1$ (dashed line) and $\alpha=0.9$ (solid line) with $\kappa=\sqrt{12}$. The weight equals to one, corresponding to relying only on current single-measurement (no memory), is shown for reference (solid horizontal line).  (b) Recovery of measured chemotactic index ($\Delta CI$)  \textcolor{black}{for live cells (solid grey circles and error bars) and simulations (black circles, dashed error bars) after a gradient-direction switch as a function of environment-to-noise ratio (ENR), defined here as the gradient squared over the  concentration. To obtain $\Delta CI$, the $CI$ of cells migrating in a stable cAMP gradient is subtracted from the $CI$ after the gradient direction is switched (see Materials).  A decaying exponential is fitted to average live-cell (solid black line) and simulated data (dashed black line). Error bars show the standard error. Arbitrary threshold separating slow and fast turns is shown by the horizontal red line. (c) Representative movement of live cells in response to low- (left) and high-ENR (right) gradient-direction switches, classified as slow and fast turns, respectively. Cell outlines in blue show the cell prior to the switch in gradient direction. Cell outlines in red show the cell after the switch. Outlines in grey are outside the time windows used for determining $CI$. Outlines are evenly spaced in time. (d) Representative cell movement from simulations in response to low- (left) and high- (right) ENR gradient-direction switches, classified as slow and fast turns, respectively. Outlines are colour-coded as in (c) and are evenly spaced in time. (Insets) Incoherent feedforward loop processes external signals before passing them to the activator of the Meinhardt model (left), characterised by  the competition between local, self-promoting activator patches via long-range inhibition (right, see Methods  for details).  (e) Percentage of turns that are fast for live-cell data (left) and simulations (right) for low and high ENRs. Distributions of turning speeds were significantly different for data (Mann-Whitney U, $p = 0.0067$) and for simulations (Mann-Whitney U, $p = 0.000096$).}}\label{Fig5}
\end{figure*}






\end{document}